\documentclass[12pt]{article}

\usepackage[dvips]{graphicx}
\usepackage{epsfig}
\usepackage{amsmath,amsfonts,amssymb,amsthm}
\usepackage{verbatim}
\usepackage{psfrag}
\usepackage{bm}
\usepackage{bbm}
\usepackage[square,comma,sort&compress,numbers]{natbib}
\usepackage{color}
\usepackage{slashed}
\usepackage{enumerate}

\usepackage{epsf,epsfig}
\usepackage{graphics}

\newcommand{\lsim}
{\;\raisebox{-.3em}{$\stackrel{\displaystyle <}{\sim}$}\;}
\newcommand{\gsim}
{\;\raisebox{-.3em}{$\stackrel{\displaystyle >}{\sim}$}\;}

\begin{document}
\thispagestyle{empty}

\begin{flushright}
{
\small
TUM-HEP-967-14
}
\end{flushright}

\vspace{0.4cm}
\begin{center}
\Large\bf\boldmath
Phenomenology of Baryogenesis from Lepton-Doublet Mixing
\unboldmath
\end{center}

\vspace{0.4cm}

\begin{center}
{Bj\"orn~Garbrecht$^1$ and Ignacio Izaguirre$^{1,2}$}\\
\vskip0.2cm
{\it $^1$Physik Department T70, James-Franck-Stra{\ss}e,\\
Technische Universit\"at M\"unchen, 85748 Garching, Germany}\\
\vskip.2cm
{\it $^2$Max-Planck-Institut f\"ur Physik (Werner-Heisenberg-Institut), F\"ohringer Ring 6,\\
80805 M\"unchen, Germany}\\
\vskip1.4cm
\end{center}

\begin{abstract}
Mixing lepton doublets of the Standard Model can lead to lepton flavour asymmetries
in the Early Universe. We present a diagrammatic representation
of this recently identified source of $CP$ violation and elaborate in detail
on the correlations between the lepton flavours
at different temperatures. For a model where two sterile right-handed neutrinos generate
the light neutrino masses through the see-saw mechanism, the lower bound
on reheat temperatures in accordance with
the observed baryon asymmetry turns out to be $\gsim 1.2\times 10^9\,{\rm GeV}$.
With three right-handed neutrinos, substantially smaller values are viable.
This requires however a tuning of the Yukawa couplings, such that there are
cancellations between the individual contributions to the masses of the light neutrinos.
\end{abstract}

%\pacs{05.30.-d, 11.30.Fs, 12.60.Fr, 14.60.St, 95.30.Cq, 98.80.-k}

%\maketitle

\section{Introduction}

Observational and theoretical studies of mixing and oscillations are typically
concerned with neutral particle states. Important examples are
neutral meson mixing,
the oscillations of Standard Model (SM) neutrinos~\cite{Beringer:1900zz}
and Leptogenesis through the mixing of sterile right-handed neutrinos (RHNs) in the
early Universe~\cite{Covi:1996wh,Flanz:1996fb,Pilaftsis:1997dr,Pilaftsis:1997jf}.
In contrast, for charged particles in the SM at vanishing temperature,
mass degeneracies between
different states are not strong enough to produce observable phenomena of
mixing and oscillations. This does however
not preclude the fact that these effects are present
in principle. Moreover, it has been demonstrated that the mixing of lepton doublets (which are gauged)
can be of importance for Leptogenesis~\cite{Endoh:2003mz,Abada:2006fw,Nardi:2006fx,Beneke:2010dz,Blanchet:2011xq}:
At high temperatures,
the asymmetries are in general  produced as superpositions of the lepton doublet flavour
eigenstates of the SM. In the SM flavour basis, this can be described in terms of
off-diagonal correlations in the two-point functions, or alternatively in
effective density-matrix formulations in terms of correlations of charge densities of
different flavours.
At smaller temperatures, interactions mediated by
SM Yukawa couplings become faster than the Hubble expansion, such that the
flavour correlations decohere. In particular, the SM leptons receive thermal mass corrections
as well as damping rates that lift the flavour degeneracy. By now, these
effects have been investigated in detail. It turns out that due to the interplay with
gauge interactions, the flavour oscillations that may be anticipated
from the thermal masses are effectively frozen, while the decoherence proceeds mainly
through the damping effects, {\it i.e.} the production and the decay of leptons in
the plasma~\cite{Beneke:2010dz,Blanchet:2011xq}.
The appropriate treatment of these flavour correlations turns out to be of leading
importance for the washout of the asymmetries from the out-of-equilibrium decays
and inverse decays of the RHNs.

The origin of the charge-parity ($CP$) asymmetry for Leptogenesis is usually
attributed to the RHNs
and their couplings~\cite{Fukugita:1986hr}.
In the standard calculation,
when describing the production and the decay of the RHNs through $S$-matrix
elements, one can diagrammatically distinguish between vertex and wave-function terms.
The presence of finite-temperature effects as well as the notorious problem of correctly
counting real intermediate states in the Boltzmann equations~\cite{Kolb:1979qa}
have motivated the use of techniques
other than the $S$-matrix approach:
It has been demonstrated that the wave-function contribution can alternatively be calculated
by solving kinetic equations (that are Kadanoff-Baym type equations
which descend from Schwinger-Dyson equations, see Refs.~\cite{Schwinger:1960qe,Keldysh:1964ud,Calzetta:1986cq,Prokopec:2003pj,Prokopec:2004ic}
on the underlying formalism)
for the RHNs and their correlations, or equivalently,
by solving for the evolution of their density matrix~\cite{De Simone:2007rw,Garny:2009qn,Garbrecht:2011aw,Garny:2011hg,Iso:2013lba,Iso:2014afa,Garbrecht:2014aga,Hohenegger:2014cpa}. The vertex contributions to the decay asymmetry can be obtained 
within the Kadanoff-Baym framework as well, as it is shown in Refs.~\cite{Buchmuller:2000nd,Garny:2009rv,Anisimov:2010aq,Garny:2010nj,Beneke:2010wd,Garbrecht:2010sz,Anisimov:2010dk}.
We note at this point
that it has more recently
been argued that the asymmetry from the wave-function correction and the contribution
from the kinetic equation are distinct contributions that should be added together~\cite{Dev:2014laa,Dev:2014wsa}. However,
it is shown in Refs.~\cite{Garny:2009qn,Garbrecht:2011aw,Garny:2011hg} that the kinetic equations derived from the
two-particle irreducible effective action capture all contributions of relevance
for the $CP$ asymmetry at
leading order, which also
encompasses the wave-function corrections.
%Adding these separately therefore leads
%to an inaccurate overcounting, {\it e.g.}
%a prediction for the asymmetry that is too large by a factor
%of two in the regime where the mass difference of the RHNs exceeds their decay widths.

The calculations for Leptogenesis based on Schwinger-Dyson equations on the
Closed-Time-Path (CTP) can also be applied to Leptogenesis from oscillations of light
(masses much below the temperature) RHNs~\cite{Drewes:2012ma},
also known as the ARS scenario after the authors~of Ref.~\cite{Akhmedov:1998qx}.
In this approach, we can interpret the $CP$ violation as originating from
cuts of the one-loop self energy of the RHNs, that are dominantly thermal.
It can be concluded that thermal effects can largely open the phase-space for $CP$-violating
cuts that are strongly suppressed for kinematic reasons at vanishing temperature.

Putting together the elements of flavour correlations for charged particles and of thermal cuts,
we can identify new sources for the lepton asymmetry, in addition to the one from
cuts in the RHN propagator. In models with multiple Higgs doublets, Higgs bosons may
be the mixing particles~\cite{Garbrecht:2012qv}, whereas in minimal type-I see-saw scenarios (with one Higgs doublet),
this role can be played by mixing SM lepton doublets~\cite{Garbrecht:2012pq}. Yet,
the RHNs remain of pivotal importance because due to their weak coupling, they
provide the deviation from thermal equilibrium that is necessary for any scenario
of baryogenesis.

While the set of free parameters of the type-I see-saw model will remain underconstrained by
present observations and those of the foreseeable future, the parameter space
in that scenario is still much smaller than in models with multiple Higgs doublets.
For our phenomenological study, we therefore choose to consider the mixing of
lepton doublets in the see-saw scenario, that is given by the Lagrangian
\begin{align}
\label{Lagrangian}
{\cal L}=&\frac{1}{2}\bar N_i({\rm i} \partial\!\!\!/\delta_{ij}-M_{Nij}) N_j
+\bar\ell_a{\rm i}\partial\!\!\!/ \ell_a
+(\partial^\mu\phi^\dagger)(\partial_\mu \phi)
\\\notag
-&Y_{ia}^*\bar\ell_a \tilde\phi P_{\rm R} N_i
-Y_{ia}\bar N_i P_{\rm L}\tilde\phi^\dagger\ell_a
-h_{ab}\phi^\dagger \bar e_{{\rm R}a} P_{\rm L} \ell_b
-h_{ab}^*\phi \bar\ell_b P_{\rm R} e_{{\rm R}_a}
\,.
\end{align}

In short, the scenario of baryogenesis from mixing lepton doublets can be described as
follows~\cite{Garbrecht:2012pq}:
Flavour-off diagonal correlations
from the mixing of active leptons $\ell_a$ (where $a$ is the flavour index)
can induce the production of lepton flavour asymmetries, corresponding to diagonal entries of a
traceless charge density matrix in flavour space. Different washout rates for the
particular flavours may then lead to a net asymmetry in total lepton number, {\it i.e,} a
non-vanishing trace of the charge density matrix. Now, since off-diagonal correlations due to mixing vanish
in thermal equilibrium, the mixing of lepton doublets that we
aim to describe consequently is  an out-of-equilibrium phenomenon.
It is thus natural to assume that initially, when the primordial plasma is close to thermal equilibrium, all correlations between the SM lepton flavours vanish. Therefore, we are
interested in possibilities of generating these dynamically. Due to gauge
interactions,
the distribution functions of the SM particles should track  their equilibrium forms very closely. Moreover,
gauge interactions are flavour-blind, so they can neither generate flavour correlations
nor destroy these (up to the indirect effects that we discuss below).
Sizeable off-diagonal correlations can however be induced through couplings to the RHNs $N$,
the distributions of which can substantially deviate from equilibrium. The flavour correlations
in the doublet leptons $\ell$ are suppressed however due to the SM Yukawa couplings $h$ with
the charged singlets $e_{\rm R}$ and the Higgs field $\phi$, where $\tilde\phi=(\epsilon\phi)^\dagger$, and
where $\epsilon$ is the totally antisymmetric ${\rm SU}(2)$ tensor. By field redefinitions,
we can impose that $h$ and $M_N$ are diagonal, which is a common and convenient choice of
basis that we adapt throughout this present paper. For simplicity, we therefore
write $M_{Ni}\equiv M_{Nii}$.

In this paper, within Section~\ref{sec:freezeout}, we first review the scenario of Ref.~\cite{Garbrecht:2012pq}.
We improve on the previous discussion by introducing a diagrammatic representation of the mechanism. Moreover,
we carefully discuss the generation and the decoherence of lepton flavour correlations at different temperatures,
paying particular attention to the fact that both effects take a finite time to fully establish.
Section~\ref{sec:surveys} contains a survey of the parameter space of baryogenesis from mixing lepton doublets
based on the Lagrangian~(\ref{Lagrangian}). Under the assumption that only two RHNs are present,
we perform a comprehensive scan, given the present best-fit values
on the light neutrino mass differences and mixing angles, such that we can identify the point in
parameter space that allows for the lowest reheat temperature for which an asymmetry in accordance
with observation can result. In addition, we show that for three RHNs, substantially smaller
temperatures can be viable, what requires however anomalously large Yukawa couplings of the $\mu$- and
the $\tau$-leptons and a cancellation in their contributions to the mass matrix of the light neutrinos.
The analysis is however restricted to the strong washout regime, such that it remains an open
question of interest whether favourable parametric regions also exist when at least one of the 
RHNs induces only a weak washout. The concluding remarks are given in Section~\ref{sec:conclusions}.

\section{Generation and Freeze-Out of the Lepton Asymmetry}
\label{sec:freezeout}

\subsection{Diagrammatic Representation of the $CP$-Violating Source Terms}

A detailed derivation of the source term for the asymmetries of
the individual lepton flavours is presented in
Ref.~\cite{Garbrecht:2012pq}, where the CTP method is employed.
Here, we do not reiterate these technical details, but we explain the qualitative
form of the main results with the help of a diagrammatic representation of the
Kadanoff-Baym equations that arise from the CTP approach.
In particular, we express the perturbative
approximations to the solutions of these equations diagrammatically.
Moreover, we discuss how the mixing of
the SM lepton doublets in the CTP formalism can be related
to a density  matrix formulation of flavour oscillations, 
that should be familiar {\it e.g.} from the problem of oscillations of
active neutrinos~\cite{Dolgov:1980cq,Barbieri:1990vx,Enqvist:1990ad,Enqvist:1991qj,Sigl:1992fn}.

\begin{figure}[t!]
\begin{center}
\begin{tabular}{c}
\epsfig{file=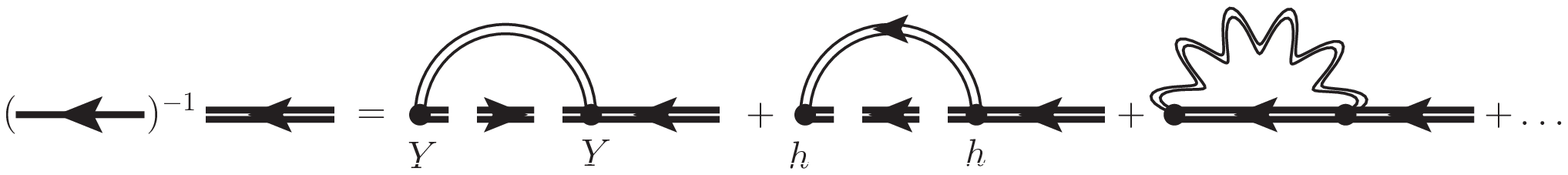,scale=0.6}\\
(A)\\
\\
\epsfig{file=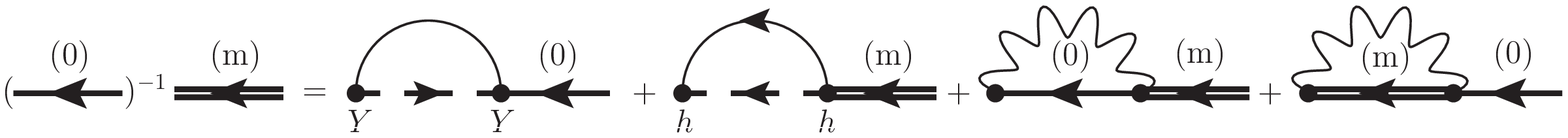,scale=0.6}\\
(B)\\\\
\epsfig{file=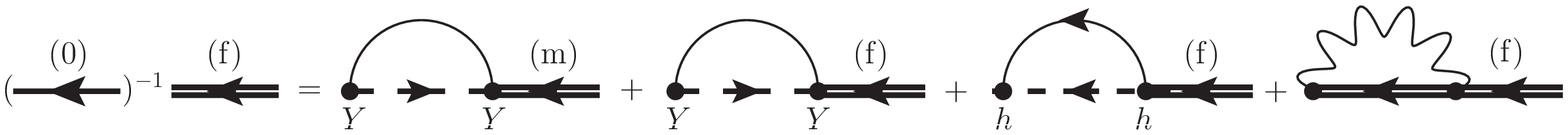,scale=0.6}
\\
(C)
\end{tabular}
\end{center}
\vskip-.4cm
\caption{
\label{fig:schwingerdyson}
The diagrams~(A) are a graphical representation of the Kadanoff-Baym equations that account
for the lepton-Yukawa interactions $h$ and $Y$ as well as gauge interactions. Single lines
stand for tree-level propagators and double lines for full propagators.
Bold solid lines with an arrow are propagators of the SM lepton doublets $\ell$, 
and dashed bold lines with an arrow of the Higgs doublet $\phi$.
Regular
solid lines with an arrow are propagators for the right-handed SM leptons $e_{\rm R}$, regular solid lines without
an arrow stand for the RHNs $N$ and wiggly lines for SM gauge bosons. The dots $\ldots$
indicate extra diagrams of different topology than those
drawn explicitly and that can be derived from the 2-particle irreducible effective action.
Figures~(B) and~(C),
illustrate the scheme that is used in Ref.~\cite{Garbrecht:2012pq} to obtain approximate solutions, which we
also apply in this work. The full propagators for the doublets $\ell$ are now
approximated by the results including flavour correlations. The loops are understood to
include gauge-mediated processes and top loops that open up the phase space for the reactions between the
particles, that are approximated to be massless, {\it cf. e.g.} Ref.~\cite{Garbrecht:2013bia}.
The superscripts $({0,\rm m},{\rm f}) $ indicate that the charge density matrix
that can be computed from the corresponding propagators for $\ell$ yields the
non-zero entries of $q_\ell^{(0,{\rm m},{\rm f})}$.
}
\end{figure}

The CTP formulation of the problem leads to Kadanoff-Baym equations, that
we show here in a diagrammatic form in Figure~\ref{fig:schwingerdyson}(A).
One may interpret the Kadanoff-Baym equations
as (a subset of) exact Schwinger-Dyson equations, that can only
be solved approximately in practice.
Since couplings can be assumed to be weak, a
perturbative one-loop expansion,
that is indicated by Figures~\ref{fig:schwingerdyson}(B) and (C), amounts to a valid approximation.

We can assume that kinetic equilibrium is established by fast gauge interactions. The distribution
functions and the propagators of the SM leptons $\ell$ are therefore effectively determined by
the matrix $q_{\ell ab}$ of the charge densities of $\ell$ and their flavour correlations~\cite{Beneke:2010dz}.
The perturbation expansion then explicitly reads
$q_\ell=q_\ell^{(0)}+q_\ell^{({\rm m})}+q_\ell^{({\rm f})}+\cdots$, where the superscript
$(\rm m)$ stands for
mixing and $({\rm f})$ for flavoured asymmetries. The zeroth order term is given by the equilibrium distribution for a vanishing charge density,
and therefore $q_\ell^{(0)}=0$. The contributions $q_\ell^{({\rm m})}$ and $q_\ell^{({\rm f})}$
are induced by the non-equilibrium right-handed neutrinos and are discussed in the following Subsections.

In order to clarify the relation of the present notation with the one used in
the derivation of Ref.~\cite{Garbrecht:2012pq}, we make the following remarks:
In the present context,
the interesting contributions within $q_\ell^{({\rm m})}$ are the
off-diagonal correlations of lepton doublets, as these are referred to in Ref.~\cite{Garbrecht:2012pq}.
The leading $CP$-violating lepton-flavour asymmetries are then contained
in $q_\ell^{({\rm f})}$.
Note also that the first term on the right-hand-side of the equation in Figure~\ref{fig:schwingerdyson}(C) is called the source term in Ref.~\cite{Garbrecht:2012pq}.

\subsubsection{Equations for Mixing Correlations}
\label{sec:flavasymm}

The first order approximation to the Kadanoff-Baym equations, that is given in
Figure~\ref{fig:schwingerdyson}(B), has the main qualitative features of
a density-matrix equation
\begin{align}
\label{density:matrix:eq}
\partial_t\varrho+{\rm i}[M,\varrho]=\Theta-\frac12\{\Gamma,\varrho\}\,,
\end{align}
where $t$ denotes time, $M$ is a mass matrix, $\Gamma$ a matrix
that describes relaxation toward equilibrium and $\Theta$ a matrix-valued inhomogeneous term.
This equation
is of a form that is familiar from many applications, among
which are neutrino
oscillations~\cite{Dolgov:1980cq,Barbieri:1990vx,Enqvist:1990ad,Enqvist:1991qj,Sigl:1992fn} and ARS Leptogenesis~\cite{Akhmedov:1998qx,Asaka:2005pn,Garbrecht:2011aw,Canetti:2012vf,Canetti:2012kh,Asaka:2011wq,Shuve:2014zua}.
In particular, it is the
commutator term $[M,\varrho]$ that induces flavour oscillations among the
off-diagonal components of $\varrho$.

Now for the present application, we replace the density matrix $\rho$
by the deviations of the lepton and anti-lepton
number densities from their equilibrium values $\delta n^{\pm({\rm m})}_{\ell}$.
The off-diagonal components of $\delta n_\ell^{\pm({\rm m})}$ describe the flavour
correlations of these non-equilibrium densities.
The matrices $\delta n^{\pm({\rm m})}_{\ell}$
then evolve according to the  following equations~\cite{Garbrecht:2012pq}:
\begin{align}
\label{eff:kin:eqs}
\frac{27\zeta(3) T_{\rm com}}{\pi^2}\partial_\eta\delta n_{\ell ab}^{\pm({\rm m})}
&\pm{\rm i}\frac{\left(h_{aa}^2-h_{bb}^2\right)T_{\rm com}^2}{16}\delta n^{\pm({\rm m})}_{\ell ab}
\\\notag
&=
-\sum\limits_i Y^\dagger_{ai}Y_{ib} B_i^Y
-(h_{aa}^2+h_{bb}^2)B_\ell^{\slashed{\rm fl}}\delta n^{\pm({\rm m})}_{\ell ab}
-B_\ell^g (\delta n^{+({\rm m})}_{\ell ab}+\delta n^{-({\rm m})}_{\ell ab})\,.
\end{align}
Here, $\eta$ is the conformal time, which is suitable for performing
calculations in the background of the expanding Universe. It is determined
up to redefinitions of the scale factor, and we make the choice that
the physical temperature in the radiation-dominated Universe is $T=1/\eta$,
what determines the comoving temperature used in above equations to be
\begin{align}
T_{\rm com}=\frac{m_{\rm Pl}}{2}\sqrt{\frac{45}{\pi^3 g_\star}}\,,
\end{align}
where $m_{\rm Pl}$ is the Planck mass and $g_\star$ the number of relativistic
degrees of freedom.
The Eqs.~(\ref{eff:kin:eqs})
are derived in Ref.~\cite{Garbrecht:2012pq} by integrating the corresponding
Kadanoff-Baym equations for distribution functions of lepton and antileptons and their
flavour correlations over the spatial momentum $\mathbf p$, and they are given diagrammatically in Figure~\ref{fig:schwingerdyson}(B).
For the distribution functions, we have assumed that gauge interactions
maintain these to be of the Fermi-Dirac
form with a matrix-valued chemical potential, as explained in Ref.~\cite{Beneke:2010dz}.
This allows to uniquely
relate the momentum-space distributions to the number densities $\delta n_\ell$
in Eq.~(\ref{eff:kin:eqs}).

We now give explaining  remarks on the individual terms in Eqs.~(\ref{eff:kin:eqs}):
\begin{itemize}
\item
The terms $\propto (h_{aa}^2-h_{bb}^2)$
are present due to thermal masses and correspond to the commutator term in the
density-matrix equation~(\ref{density:matrix:eq}). Notice the different sign
of these terms in the equations for lepton and anti-lepton densities,
that was first noted in Ref.~\cite{Beneke:2010dz}. This is what makes it necessary
to treat the lepton and anti-lepton densities differently, rather than considering
the matrix of charge densities and their and correlations $q_\ell^{({\rm m})}=\delta n_\ell^{+({\rm m})}-\delta n_\ell^{-({\rm m})}$.
\item
The terms involving $B_i^Y$
describe the decays and inverse decays of sterile neutrinos into the active
leptons. Provided the distribution of the sterile neutrinos $N$ deviates from
thermal equilibrium, these processes induce the flavour correlations of active leptons
through the off-diagonal
components ($a\not=b$) of $Y^\dagger_{ai}Y_{ib}$ in first place.
In this work,
we restrict ourselves to the parametric regime where the freeze-out value
of the lepton asymmetry is determined at times when the the distribution
of the $N_i$ is dominated by
non-relativistic particles, commonly referred to as strong washout.
For this situation, we can assume that $M_{Ni}\gg T$,
and approximate the rate for decays and inverse decays using~\cite{Garbrecht:2012pq}
\begin{align}
\label{B:Y}
B^Y_i\approx
-\frac{T^{\frac32}M_{Ni}^\frac72}{2^\frac{13}{2}\pi^\frac52}
\frac{\mu_{Ni}}{T}{\rm e}^{-\frac{M_{Ni}}{T}}\,.
\end{align}
Here, $\mu_{Ni}$ denotes a pseudochemical potential that can be employed in
order to describe the deviation
\begin{align}
\label{deltafN:mu}
\delta f_{Ni}(\mathbf p)=f_{Ni}(\mathbf p)-f_{Ni}^{\rm eq}(\mathbf p)
\approx\frac{\mu_{Ni}}{T}{\rm e}^{\sqrt{\mathbf{p}^2 + M_{Ni}^2}/T}
\end{align}
of  $f_{Ni}$, which is the distribution function of the
sterile neutrino $N_i$, from its equilibrium
form $f_{Ni}^{\rm eq}$. For standard Leptogenesis, results for the lepton asymmetry obtained when using a distribution function numerically derived from the Boltzmann equations before momentum-averaging are compared to the results obtained when using the approximation with a pseudo-chemical potential in Refs.~\cite{Beneke:2010wd,Basboll:2006yx}, and the discrepancy between the two methods
is found to be small because for strong washout, the RHNs are non-relativistic, such that
they mostly populate modes with momenta that are small compared to the temperature. Within the density-matrix equation~(\ref{density:matrix:eq}), the
$B^Y_i$ terms correspond to the inhomogeneous term $\Theta$. In the diagrammatic equation
Figure~\ref{fig:schwingerdyson}(B), this is the first term on the right-hand side.
\item
The terms with $B_\ell^{\slashed{\rm fl}}$ describe the decay of the flavour
correlations due to the SM Yukawa interactions, that discriminate
between the different lepton flavours. In the
density-matrix equation~(\ref{density:matrix:eq}), they correspond to the anticommutator
term involving the relaxation rate $\Gamma$,
and in Figure~\ref{fig:schwingerdyson}(B) to the second term on the right-hand side.
The relevant processes involve the radiation
of extra gauge bosons or the decay and inverse decay of a virtual Higgs
boson into a pair of top quarks, which are understood to be contained
in the loops. (Otherwise, the $1\leftrightarrow2$ processes between
approximately massles particles would be strongly suppressed kinematically.) The rates for these processes are calculated
to LO in Ref.~\cite{Garbrecht:2013bia}, where it is found that
$\gamma^{\rm fl}=5\times10^{-3}$
(see also Ref.~\cite{Cline:1993bd} for an earlier estimate that
leads to a similar quantitative conclusion and
Refs.~\cite{Anisimov:2010gy,Besak:2012qm} for
a recent LO calculation for the production of massless sterile neutrinos, that
is closely related). Taking the momentum average of the kinetic
equations of Ref.~\cite{Garbrecht:2012pq}, we find that
$B_\ell^{\slashed{\rm fl}}=54\zeta(3)/\pi^2\times\gamma^{\rm fl} T \approx3.3\times10^{-2} T^2$,
what differs from the value used in Ref.~\cite{Garbrecht:2012pq} due to
the updated result of Ref.~\cite{Garbrecht:2013bia}. We note that the averaging
implies the assumption that flavour correlations in all momentum modes
decay at the same rate, which is not the case in reality. This procedure should
therefore incur an order one inaccuracy that may be removed in the future by extra numerical
efforts.
\item
Finally, the contribution with $B_\ell^g$ describes pair creation and annihilation
processes that drive $\delta n^+_{ab}+\delta n^-_{ab}$ toward zero.
It thus forces an alignment between the correlations among the different
flavours of leptons and of anti-leptons, that would otherwise perform
oscillations with opposite angular frequency, and as a consequence, the
evolution of the off-diagonal correlations is overdamped~\cite{Beneke:2010dz,Blanchet:2011xq}.
Gauge interactions thus contribute indirectly to the decay of flavour correlations
in addition to the direct damping through the Yukawa interactions.
In Ref.~\cite{Garbrecht:2012pq}, the relevant momentum average of the
pair creation and annihilation rate is estimated as
$B_\ell^g=1.7\times 10^{-3} T^2$, based on the thermal rates for $s$-channel mediated
processes, that should yield the dominant contribution due to the
large number of degrees of freedom in the SM.
\end{itemize}

Ignoring the derivative with respect to the conformal time
$\eta$, the solution to Eqs.~(\ref{eff:kin:eqs})
is given by
\begin{align}
\label{q:ab}
q^{({\rm m})}_{\ell ab}=
\delta n_{\ell ab}^{+({\rm m})}-\delta n_{\ell ab}^{-({\rm m})}
&=
{\rm i}\Xi_{ab}
({\cal Q}_{\ell ab}/T^2)\sum_i Y_{ai}^\dagger Y_{ib} B_i^Y
\,,
\end{align}
where
\begin{align}
\label{Qellab}
{\cal Q}_{\ell ab}=\frac
{(h_{aa}^2-h_{bb}^2)(T^4/8)}
{\left[(h_{aa}^2-h_{bb}^2)/16\right]^2 T^4+(h_{aa}^2+h_{bb}^2)B^{\slashed{\rm fl}}[2 B_\ell^g+(h_{aa}^2+h_{bb}^2)B_\ell^{\slashed{\rm fl}}]}
\end{align}
and where $\Xi$ is a matrix specified in Eq.~(\ref{matrix:Xi}) below, that selects only
those correlations that have enough time to be built up at the temperatures of interest.
The flavour matrix ${\cal Q}_{\ell}$ can be viewed as the quantity that multiplies the $CP$-violation originating
from the Yukawa couplings $Y$ of the sterile neutrinos. By comparing the powers of
$h$ in the numerator and the denominator, we explicitly see that
the $q^{({\rm m})}_{\ell ab}$ become suppressed if $h_{aa}$ or $h_{bb}$ become large. This
is because processes mediated the SM lepton Yukawa interactions lead to decoherence these
off-diagonal correlations, as it is familiar from flavoured
Leptogenesis~\cite{Abada:2006fw,Nardi:2006fx,Beneke:2010dz}.
Note that these decoherence effects also avoid the resonance catastrophe that would occur
for $h_{aa}\to h_{bb}$ in their absence.

Neglecting the time derivatives is justified provided that there
is enough time for the flavour correlations to adapt to the change in
the non-equilibrium density of the sterile neutrinos. From Eqs.~(\ref{eff:kin:eqs}), 
it can be seen that rate for the flavour correlations to build up is given by
\begin{align}
\label{gammabuildup}
\Gamma^{-}_{q_{\ell ab}}
=\frac{\pi^2}{54\zeta(3) T}\left(
B_\ell^g
+(h_{aa}^2+h_{bb}^2)B_\ell^{\slashed{\rm fl}}
-\sqrt{{B_\ell^g}^2-\left[(h_{aa}^2-h_{bb}^2)T^2/16\right]^2}
\right)\,.
\end{align}
This implies that the correlations do not build up in case the entries of $h$ are so small
that $\Gamma^{-}_{q_{\ell ab}}\lsim H$ when the right-handed neutrinos go out of equilibrium.
One should expect this, because in the limit
$h\to 0$, the system is flavour blind and there should be no dynamical generation
of correlations. For the subsequent discussion, it is useful to
compare the rate for the build up of correlations~(\ref{gammabuildup})
with the more commonly employed rate of flavour equilibration
\begin{align}
\Gamma^{-}_{q_{\ell xy}}\approx \pi^2{54\zeta(3)}B_\ell^{\slashed{\rm fl}} |h_{xx}|^2T\approx0.15\times B_\ell^{\slashed{\rm fl}} |h_{xx}|^2 T\,,
\end{align}
which is valid for $|h_{xx}|\gg |h_{yy}|$ and for the size of the particular lepton
Yukawa couplings as in the SM.

\subsubsection{Equations for the Flavoured Asymmetries}
\label{sec:regimes}

In deriving Eq.~(\ref{Qellab}), we have accurately taken account of the impact
of the gauge and the Yukawa couplings on generating and also damping off-diagonal correlations
in $q_\ell^{(\rm m)}$. In the equation represented by Figure~\ref{fig:schwingerdyson}(C), the same
couplings enter once more. At this level, we adapt
from the usual calculations on flavoured Leptogenesis~\cite{Abada:2006fw,Nardi:2006fx}
the simplifying approximation that flavour
correlations in $q_{\ell ab}^{({\rm f})}$ are either unaffected (unflavoured regime) or completely
erased (fully flavoured regime). We note that a detailed calculation as for $q_\ell^{(\rm m)}$ should
have the result that in the fully flavoured regime, the according components
of $q_{\ell ab}^{({\rm f})}$ are suppressed rather than fully erased, what would lead to
sub-leading corrections to the present calculations.
These flavour effects suggest
to distinguish between the following regimes:
\begin{enumerate}[\label=A]
\item
\label{regime:2flavour}
\begin{enumerate}[\label=1]
\item
\label{regime:2flavour:high}
When $2.7\times10^{11}\, {\rm GeV} \gsim T\gsim 4.1\times10^{10}\, {\rm GeV}$, off-diagonal
correlations in $q_\ell^{({\rm m})}$ will not have enough time to build up, since
$\Gamma^{-}_{q_{\ell \tau \mu}},\Gamma^{-}_{q_{\ell \tau e}}<H$. Leptogenesis
from mixing lepton doublets should therefore be inefficient at these temperatures.
\item
\label{regime:2flavour:low}
For $4.1\times10^{10}\, {\rm GeV} \gsim T\gsim 1.3\times10^9\, {\rm GeV}$,
the correlations involving $\tau$ will build up in $q_\ell^{(\rm m)}$, but
$q^{(\rm m)}_{\ell \mu e}=q^{(\rm m)}_{\ell e \mu}\approx 0$  due to
$\Gamma^{-}_{q_{\ell \mu e}}\ll H$. Within $q_\ell^{(\rm f)}$, the correlations involving
$\tau$ are erased due to decohering scattering. In summary, the non-zero entries
of the charge-density matrices are given by
\begin{align}
q_\ell^{(\rm m)}=
\left(
\begin{array}{ccc}
0 & 0 & q^{({\rm m})}_{\ell e\tau}\\
0 & 0 & q^{({\rm m})}_{\ell \mu\tau}\\
q^{({\rm m})*}_{\ell e\tau} &  q^{({\rm m})*}_{\ell \mu\tau} & 0
\end{array}
\right)\,,
\quad
q_\ell^{(\rm f)}=
\left(
\begin{array}{ccc}
q^{({\rm f})}_{\ell ee} & q^{({\rm f})}_{\ell e\mu} & 0\\
q^{({\rm f})*}_{\ell e\mu} & q^{({\rm f})}_{\ell \mu\mu} &0\\
0 & 0 & q^{({\rm f})}_{\ell \tau\tau}
\end{array}
\right)\,.
\end{align}
\end{enumerate}
\item
\label{regime:3flavour}
\begin{enumerate}[\label=1]
\item
\label{regime:3flavour:high}
When $1.3\times 10^9\, {\rm GeV}\gsim T\gsim 2.0\times 10^8\, {\rm GeV}$,
the correlations $q^{(\rm m)}_{\ell e\tau}$ and $q^{(\rm m)}_{\ell\mu\tau}$
have enough time to build up, but still not $q^{({\rm m})}_{\ell e\mu}$.
All off-diagonal 
correlations in $q_\ell^{(\rm f)}$
decay and should be set to zero due to the direct damping of flavour correlations:
\begin{align}
q_\ell^{(\rm m)}=
\left(
\begin{array}{ccc}
0 &0 & q^{({\rm m})}_{\ell e\tau}\\
0 & 0 & q^{({\rm m})}_{\ell \mu\tau}\\
q^{({\rm m})*}_{\ell e\tau} &  q^{({\rm m})*}_{\ell \mu\tau} & 0
\end{array}
\right)\,,
\quad
q_\ell^{(\rm f)}=
\left(
\begin{array}{ccc}
q^{({\rm f})}_{\ell ee} & 0 & 0\\
0 & q^{({\rm f})}_{\ell \mu\mu} &0\\
0 & 0 & q^{({\rm f})}_{\ell \tau\tau}
\end{array}
\right)\,.
\end{align}
\item
\label{regime:3flavour:low}
When $T\lsim 2.0\times 10^8\, {\rm GeV}$, all correlations in $q_\ell^{(\rm m)}$ have
enough time to build up and all off-diagonal correlations in $q_\ell^{(\rm f)}$ are
erased, such that
\begin{align}
q_\ell^{(\rm m)}=
\left(
\begin{array}{ccc}
0 & q^{({\rm m})}_{\ell e\mu} & q^{({\rm m})}_{\ell e\tau}\\
q^{({\rm m})*}_{\ell e\mu} & 0 & q^{({\rm m})}_{\ell \mu\tau}\\
q^{({\rm m})*}_{\ell e\tau} &  q^{({\rm m})*}_{\ell \mu\tau} & 0
\end{array}
\right)\,,
\quad
q_\ell^{(\rm f)}=
\left(
\begin{array}{ccc}
q^{({\rm f})}_{\ell ee} & 0 & 0\\
0 & q^{({\rm f})}_{\ell \mu\mu} &0\\
0 & 0 & q^{({\rm f})}_{\ell \tau\tau}
\end{array}
\right)\,.
\end{align}
\end{enumerate}
\end{enumerate}
When above constraints on the rates for the creation and for the decay
of flavour correlations 
are not sufficiently saturated, a treatment of incomplete flavour decoherence is in order. This
has been put forward in Ref.~\cite{Beneke:2010dz}, but it is yet numerically challenging and
requires further developments. For the numerical examples presented in this paper,
it turns out however that the approximations as described for regime~\ref{regime:3flavour:high} should be appropriate.

In view of these considerations for the generation of flavour correlations,
we can now write down the expressions for the matrix $\Xi$ introduced in Eq.~(\ref{q:ab}),
that depend on the temperature regime in which Leptogenesis takes place:
\begin{align}
\label{matrix:Xi}
\Xi=
\left(
\begin{array}{ccc}
1 & 0 & 1\\
0 & 1 & 1\\
1 & 1 & 1\\
\end{array}
\right)
\,\;\textnormal{in Regimes~\ref{regime:2flavour:low} and~\ref{regime:3flavour:high}}\,,\quad
\Xi=
\left(
\begin{array}{ccc}
1 & 1 & 1\\
1 & 1 & 1\\
1 & 1 & 1\\
\end{array}
\right)
\,\;\textnormal{in Regime~\ref{regime:3flavour:low}}\,.
\end{align}
Note that the diagonal components are actually irrelevant, because ${\cal Q}_{\ell aa}=0$.

From above explicit expressions that indicate the non-zero entries of
$q_\ell^{({\rm m})}$ and $q_\ell^{({\rm f})}$, we see that these charge-density
matrices are complementary, what justifies the decomposition of the Kadanoff-Baym
equations done in Ref.~\cite{Garbrecht:2012pq}, which is represented here in the
Figures~\ref{fig:schwingerdyson}(B) and~(C). We should still make a remark though on the
fact why we do not consider terms of order $Y^2$ that multiply
$q_\ell^{({\rm m,f})}$ on the right-hand side of Figure~\ref{fig:schwingerdyson}(B).
The reason is that by virtue of the requirement $\Gamma^-_{q\ell ab}>H$ for the
non-zero correlations $q_{\ell ab}^{({\rm m})}$, the
relation $(h_{aa}^2+h_{bb}^2)\gamma^{\rm fl}>H$ should be amply fulfilled as well.
Therefore, the flavour damping mediated by the SM Yukawa couplings is more efficient in
suppressing the off-diagonal correlations in $q_{\ell ab}^{({\rm m})}$ than the damping induced
by the couplings $Y$, that we therefore neglect. Note however, that in the equation
represented by Figure~\ref{fig:schwingerdyson}(C), rates of order  $Y^2$ multiply
$q_\ell^{({\rm f})}$ and $q_\ell^{({\rm m})}$. In particular, the second term on the right
hand side is the washout term that suppresses the diagonal charge densities and those of 
the off-diagonal
components, that are unaffected by flavour effects. The second term on the
right-hand side is the source term, that we discuss next.

The charge correlations $q_\ell^{({\rm m})}$, as given by Eq.~(\ref{q:ab}),
are of an out-of-equilibrium form,
which can simply be inferred from the fact that they are purely off-diagonal.
Therefore, they give rise to a non-vanishing source of correlations for
entries of the matrix $q_\ell^{({\rm f})}$, which may be diagonal or non-diagonal.
The first term on the right-hand side of Figure~\ref{fig:schwingerdyson}(C) is the
$CP$ violating source that creates flavour asymmetries at the rate
\begin{align}
\label{source:q}
\partial_\eta q^{({\rm f})}_{\ell ab}
&=\sum_i\frac{3M_{Ni}^{5/2}T^{1/2}}{2^{9/2}\pi^{5/2}}{\rm e}^{-\frac{M_{Ni}}{T}}
\left(Y^\dagger_{ai} Y_{ic} q^{({\rm m})}_{\ell cb}+q^{({\rm m})}_{\ell ac} Y^\dagger_{ci}Y_{ib}\right)
\,.
\end{align}
Note that the flavour structure of this equation takes the anticommutator form
present in Eq.~(\ref{density:matrix:eq}).
When compared with the corresponding result of
Ref.~\cite{Garbrecht:2012pq}, we have generalised this source of
lepton flavour asymmetries
such that
it also includes the off-diagonal correlations that are generated and that should be relevant
for the regime~\ref{regime:2flavour:low}.

\subsubsection{Comparision with the Source Term for Conventional Leptogenesis}

It is of course of interest to compare the asymmetry from lepton mixing with the
standard asymmetry from the decays and the mixings of sterile neutrinos.
For this purpose, we make use of the standard decay asymmetry including flavour correlations, that is given by~\cite{Covi:1996wh,Beneke:2010dz,Antusch:2011nz}
\begin{align}
\label{decayasymmetry:standard}
\varepsilon^{Ni}_{ab}=
-\frac{3}{16\pi [Y Y^\dagger]_{ii}}\sum\limits_{j\not=i}
\Big\{
{\rm Im}\left[Y^\dagger_{ai}(Y^*Y^t)_{ij}Y_{jb}\right]\frac{\xi(x_j)}{\sqrt{x_j}}
+{\rm Im}\left[Y^\dagger_{ai}(YY^\dagger)_{ij}Y_{jb}\right]\frac{2}{3(x_j-1)}
\Big\}\,,
\end{align}
where $x_j=(M_{Nj}/M_{Ni})^2$ and
\begin{align}
\xi(x)=\frac23 x\left[(1+x)\log\frac{1+x}{x}-\frac{2-x}{1-x}\right]
\,.
\end{align}

We should compare the function $\xi(x)$ with ${\cal Q}_{\ell ab}$ given
in Eq.~(\ref{Qellab}), since both play the role of loop factors, that
multiply the $CP$ asymmetry that is present in the Yukawa couplings
$Y$. Inspecting ${\cal Q}_{\ell ab}$ and ignoring first those of the denominator terms
that involve $B^{\slashed{\rm fl}}$, we observe an enhancement
$\sim1/(h_{aa}^2-h_{bb}^2)$, which one might naively guess when expecting that
the source from lepton mixing is enhanced by the difference of the square of
the thermal masses of the leptons. Within $\xi(x_j)$, the corresponding
enhancement is explicitly
present in the terms $\sim1/(M_{Ni}^2-M_{Nj}^2)$. However, within ${\cal Q}_{\ell ab}$,
the denominator terms that involve $B^{\slashed{\rm fl}}$ indicate that this enhancement
is limited by the damping of the flavour correlations, which induce $CP$ violation from
mixing. We observe two different types of damping: First, the terms
$\sim B^{\slashed{\rm fl}}$ are due to the decoherence of correlations due to
scatterings mediated by the SM-lepton Yukawa couplings $h$. Second, the terms
$\sim B^{\slashed{\rm fl}}B^g$ originate from the effect that leptons and anti-leptons
oscillate with opposite frequencies. In conjunction with pair creation and annihilation
processes, that mediate between leptons and anti-leptons, this leads to flavour decoherence
from overdamped oscillations, as it was first described in Ref.~\cite{Beneke:2010dz} and
as it is confirmed in Ref.~\cite{Blanchet:2011xq}.
An appropriate treatment of the resonant limit  $x\to 1$ reveals
a similar regulating behaviour also for standard Leptogenesis~\cite{Garbrecht:2011aw,Garny:2011hg,Iso:2013lba,Iso:2014afa,Garbrecht:2014aga}.

\subsection{Flavour Correlations and Spectator Effects}

When Leptogenesis occurs at high temperatures, where flavour effects
are not important, and when the production and the washout of the
asymmetry results from decays and inverse decays of the lightest
sterile neutrino, which we call $N_1$ for now in order to be definite, it
is convenient to perform the single-flavour or vanilla approximation.
It is based on a unitary flavour transformation (of the left-handed SM
leptons) such that $N_1$ only couples to one linear combination
$\ell_\parallel$ of the left-handed leptons.
On the other hand, when Leptogenesis occurs at temperatures below
$1.3\times10^9\,{\rm GeV}$, it is advantageous to remain in the basis where the
SM Yukawa-couplings $h$ are diagonal. Interactions mediated by
these couplings then rapidly destroy off-diagonal flavour correlations.
In a practical calculation, we may then just delete the off-diagonal
correlations that are induced by the terms~(\ref{source:q})
and~(\ref{decayasymmetry:standard})~\cite{Abada:2006fw,Nardi:2006fx}.
In the range between $1.3\times10^9\,{\rm GeV}$ and $2.7\times10^{11}\,{\rm GeV}$, it is often convenient to
perform a two flavour-approximation, where lepton asymmetries
are deposited in the flavour $\tau$ and in a linear combination $\sigma$ of
$e$ and $\mu$. Correlations between $\tau$ and $\sigma$ are then erased
by $h_{\tau\tau}$-mediated interactions. In effect, only diagonal
correlations in the flavours $\tau$ and $\sigma$ need to be calculated.

However the reduction to a single flavour at high temperatures or to two uncorrelated
flavours
between $1.3\times10^9\,{\rm GeV}$ and $2.7\times10^{11}\,{\rm GeV}$ only works when
$N_1$ is the only right-handed neutrino that is effectively produced or destroyed in decay and inverse
decay processes at times relevant for Leptogenesis,
which is typical for hierarchical scenarios where
$M_1\ll M_{2,3}$, and therefore, the heavier of the $N_i$ are strongly Maxwell suppressed.
Once more than one of the right-handed neutrinos is effectively produced or destroyed, flavour
correlations of the active leptons must be taken into
account~\cite{Blanchet:2011xq,Antusch:2010ms}, because the combinations $\ell_\parallel$
or $\sigma$ are in general different for the individual $N_i$.
Now, because for Leptogenesis
from mixing lepton doublets, the relevant
$CP$-violating cut is purely thermal and it involves a right-handed neutrino
that must be different from the decaying neutrino, we must require that
there are at least two sterile neutrinos $N_{1,2}$ with masses $M_{N1,2}$ that
are not hierarchical. The latter condition is imposed in order to avoid
a Maxwell suppression of the $CP$ asymmetry\footnote{This can be seen when substituting
Eq.~(\ref{B:Y}) into Eq.~(\ref{source:q}) and noting that the non-vanishing terms always
involve two different $N_i$ along with their distribution functions.}. This implies that for
regime~\ref{regime:2flavour:low}, we must account for lepton flavour correlations,
which in general cannot be avoided by a basis transformation due to the dynamical
importance of two sterile neutrinos. In regime~\ref{regime:3flavour}, we may proceed
by remaining in the basis where the SM-lepton Yukawa-couplings $h$ are diagonal and
simply delete the off-diagonal correlations in the source
terms~(\ref{source:q}) and~(\ref{decayasymmetry:standard}).

In the CTP approach, the collision terms including flavour correlations
can be derived in a systematic and straightforward manner. The flavour
structure turns out to be in accordance with the anticommutator term
in the density matrix equation~(\ref{density:matrix:eq}). In order to
further improve on the accuracy of the present analysis compared to the
one presented in Ref.~\cite{Garbrecht:2012pq}, we include the partial
redistribution and equilibration of the flavoured asymmetry within the
SM particles present in the plasma, the so-called spectator
effects~\cite{Barbieri:1999ma,Buchmuller:2001sr,Davidson:2008bu}.
The relevant processes are mediated by Yukawa interactions as well
as by the strong and weak sphalerons.
It is then useful to track within the Boltzmann equations
those asymmetries and correlations for which the diagonal parts are only violated by the
decays and the inverse decays of the sterile neutrinos. These are given
by
\begin{subequations}
\begin{align}
\Delta_{aa}=&B/3-L_a\,,\\
\label{Delta:q:offdiag}
\Delta_{ab}=&-2 q^{({\rm f})}_{\ell ab}\quad \textnormal{for}\;\;a\not=b\,,
\end{align}
\end{subequations}
which we have formulated as a matrix-valued quantity in view
of the Boltzmann equations including flavour coherence,
that we formulate below. Here, the number density of baryons is given
by $B$, and the diagonal number density of leptons of the flavour $a$ by $L_a$,
{\it i.e.} it accounts for left- and right-handed SM leptons.
In addition to the
interactions with sterile neutrinos, the flavour correlations
in $q_\ell^{({\rm f})}$ are also
altered by processes that are mediated by
the SM-lepton Yukawa interactions. However, according to our above
discussion, we assume that these are either negligible (unflavoured
regime) or lead to
a complete decoherence (fully flavoured regime). We also note
that the quantities denoted by $q_X$ are defined here as the charge
densities within a single component of the gauge multiplet $X$. In contrast,
$\Delta$ accounts for a the total charge density that is summed over
all gauge multiplicities. This implies that if $q_{\ell aa}$ changes
by two units, $\Delta_{aa}$ does so by minus one.

In order to obtain the washout rates, we must reexpress the $q_{\ell aa}$
in terms of the $\Delta_{aa}$. Moreover, there is also an asymmetry in Higgs
bosons that depends on the $\Delta_{aa}$. We obtain these densities through
the relations
\begin{subequations}
\label{A:C}
\begin{align}
q_{\ell aa}=&-\frac12 \sum\limits_{b\in\{e,\mu,\tau\}}A_{ab} \Delta_{bb}\,,\\
q_\phi=&\frac 12 \sum\limits_{b\in\{e,\mu,\tau\}}C_{\phi b} \Delta_{bb}\,.
\end{align}
\end{subequations}
As explained above, the redistribution of the asymmetries due to the spectators
only afflicts the diagonal components of $q_{\ell}$.
For regime~\ref{regime:2flavour}, we
take strong sphalerons, weak sphalerons (that couple to the trace of $q_\ell$),
interactions mediated by Yukawa couplings of the $t,b,c$ quarks and
$\tau$ leptons to be in equilibrium. This leads to
\begin{subequations}
\begin{align}
A=&\frac{1}{589}
\left(
\begin{array}{ccc}
-503 & 86 & 60 \\
86 & -503 & 60 \\
30 & 30 & -390
\end{array}
\right)\,,\\
C_\phi=&
-\frac{1}{589}
\left(
\begin{array}{ccc}
164 & 164 & 216
\end{array}
\right)\,.
\end{align}
\end{subequations}
In regime~\ref{regime:3flavour}, the $\tau$-lepton and $s$-quark mediated
interactions equilibrate in addition, what leaves us with
\begin{subequations}
\begin{align}
A=&\frac{1}{1074}
\left(
\begin{array}{ccc}
-906 & 120 & 120 \\
75 & -688 & 28 \\
75 & 28 & -688
\end{array}
\right)\,,\\
C_\phi=&
-\frac{1}{179}
\left(
\begin{array}{ccc}
37 & 52 & 52
\end{array}
\right)\,.
\end{align}
\end{subequations}
Note that the
factors of $1/2$ in Eqs.~(\ref{A:C}) are due to ${\rm SU}(2)$ doublet nature
of $\ell$ and $\phi$ and to our convention of $q_{\ell aa}$ and $q_\phi$ to account
for one component of the ${\rm SU}(2)$ doublet only.

\subsection{Boltzmann Equations}
\label{sec:Boltzmann}

With above explanations and remarks, and with the calculational details given
in Ref.~\cite{Garbrecht:2012pq}, we put together Boltzmann equations that
describe the freeze-out of the lepton asymmetry. In contrast to the most
commonly studied scenarios, we now have more than one sterile neutrino in
the game. Therefore, we distinguish the asymmetries that are created through
the decays and inverse decays of the individual $N_i$ as $q^{({\rm m,f})Ni}_{\ell}$
and $\Delta^{Ni}$, such that Eq.~(\ref{q:ab}) is decomposed as
\begin{align}
q^{({\rm m})Ni}_{\ell ab}=\frac{{\cal Q}_{ab}}{T^2}Y^\dagger_{ai}Y_{ib} B_i^Y\,.
\end{align}
Furthermore, we follow the common procedure
of expressing the Boltzmann equations in terms of the ratios $Y_{Ni}=n_{Ni}/s$,
$Y^{Ni}_\ell=q^{({\rm f})Ni}_\ell/s$, $Y^{Ni}_\phi=q_\phi^{Ni}/s$ and
$Y^{Ni}_\Delta=\Delta^{Ni}/s$, where $s$ denotes the entropy density.

It is convenient to parametrise the time evolution through  variables
$z_i=M_{Ni}/T$, in terms of which the equations that describe the freeze-out
of the asymmetry can be expressed
in the following approximate form:
\begin{subequations}
\label{kineq:strongwashout}
\begin{align}
\frac{dY^{Ni}_{\Delta}}{dz_i}&=-2 \bar S_{\ell}^{Ni} (Y_{Ni}-Y_{Ni}^{\rm eq})
-\bar W_\Delta[Y^{Ni}_{\Delta}]
\,,
\\
\frac{dY_{Nk}}{dz_i}&=\bar{\cal C}_{Nk}(Y_{Nk}-Y_{Nk}^{\rm eq})\,,
\end{align}
\end{subequations}
where $Y_{Ni}^{\rm eq}$ is the equilibrium value of $Y_{Ni}$.
Through Eqs.~(\ref{B:Y},\ref{deltafN:mu},\ref{q:ab},\ref{source:q}), we can identify
the source term for Leptogenesis from mixing leptons
\begin{align}
\label{Sbar}
\bar S_{\ell ab}^{Ni}&=
\Lambda_{ab}
\sum\limits_{\underset{j\not=i}{jc}}
\frac{3 a_{\rm R}z_i^{\frac92}{\rm e}^{-\frac{M_{Nj}}{M_{Ni}}z_i}}{2^{23/2}\pi^{7/2}}
\frac{M_{Nj}^\frac72}{M_{Ni}^\frac92}
\frac{[YY^\dagger]_{ii}}{[YY^\dagger]_{jj}}
{\rm i}
\left(
{\cal Q}_{\ell cb}Y^\dagger_{ai}Y_{ic}Y^\dagger_{cj}Y_{ja}
+{\cal Q}_{\ell ac}Y^\dagger_{aj}Y_{jc}Y^\dagger_{ci}Y_{ia}
\right)
\,.
\end{align}
In the diagrammatic representation, this source corresponds to the first term
on the right-hand side of Figure~\ref{fig:schwingerdyson}(C).
According to what we state above regarding the effect of the
SM lepton-Yukawa couplings, we should choose
\begin{subequations}
\begin{align}
\Lambda=
\left(
\begin{array}{ccc}
\Lambda_{ee} & \Lambda_{e\mu} & \Lambda_{e\tau}\\
\Lambda_{\mu e} &\Lambda_{\mu\mu} & \Lambda_{\mu\tau}\\
\Lambda_{\tau e} & \Lambda_{\tau\mu} &\Lambda_{\tau\tau}
\end{array}
\right)
=&
\left(
\begin{array}{ccc}
1 & 1 & 0\\
1 & 1 & 0\\
0 & 0 & 1
\end{array}
\right)\,\quad\textnormal{in Regime~\ref{regime:2flavour}}\,,\\
\Lambda=&
\left(
\begin{array}{ccc}
1 & 0 & 0\\
0 & 1 & 0\\
0 & 0 & 1
\end{array}
\right)\,\quad\textnormal{in Regime~\ref{regime:3flavour}}\,,
\end{align}
\end{subequations}
such that correlations that are suppressed by Yukawa-mediated interactions
faster than the Hubble rate are deleted from the outset. We re-emphasise that
a transformation to an effective two-flavour basis is not
possible in Regime~\ref{regime:2flavour} when more than one
of the sterile neutrinos is involved in the washout~\cite{Blanchet:2011xq,Antusch:2010ms}.

The Boltzmann equations~(\ref{kineq:strongwashout}) apply to standard Leptogenesis
as well. In that case, the source is given by
\begin{align}
\label{Sbar:standard}
\bar S_{\ell ab}^{Ni}&=\Lambda_{ab}\frac12 \bar {\cal C}_{Ni} \varepsilon^{Ni}_{ab}\,.
\end{align}
We have defined here the $\bar S_{\ell ab}$ in Eqs.~(\ref{Sbar},\ref{Sbar:standard})
such that these comply for $a=b$ with the expressions given in
Ref.~\cite{Garbrecht:2012pq}, where these are introduced as a source for
$q_\ell$ rather than $\Delta$. The various factors $-1/2$ and $-2$ 
therefore account for the $\ell$ being ${\rm SU}(2)$ doublets.

The decay rate of the sterile neutrino $N_k$ in terms of the variable $z_i$ is
\begin{align}
\label{Cbar:N}
\bar {\cal C}_{N k}=\frac{1}{8\pi}\sum\limits_{a}Y_{ka}Y^\dagger_{ak}a_{\rm R}z_i\frac{M_{Nk}}{M_{Ni}^2}\,.
\end{align}
In the non-relativistic approximation, this agrees with its thermal average up to
relative corrections of order $T/M_{Nk}$.
In the strong washout
regime, a substantial simplification arises from the fact
that the deviation of the sterile neutrinos
from equilibrium is small, such that we can approximate
\begin{align}
(Y_{Nk}-Y_{Nk}^{\rm eq})\approx\frac{1}{\bar{\cal C}_{Nk}}\frac{dY^{\rm eq}_{Nk}}{dz_i}\,.
\end{align}
We use this relation for both scenarios, Leptogenesis from mixing leptons as well
as from the decay and mixing of sterile neutrinos. The error incurred through
this standard approximation is investigated in Refs.~\cite{Beneke:2010wd,Basboll:2006yx}.

Finally, we need to obtain an expression for the washout rate
$\bar W[Y_\ell^{Ni}]$, that is of the anticommutator form indicated in the
density matrix equation~(\ref{density:matrix:eq}) and that should account for
the spectator effects. In terms of Feynman diagrams, the washout
term corresponds to the second graph on the right-hand side of Figure~\ref{fig:schwingerdyson}(C).
As explained above, washout affects the flavour-diagonal
lepton charges and the off diagonal correlations in a different manner, such that
it is useful to define
\begin{subequations}
\begin{align}
Y_\Delta^{\rm diag}=&{\rm diag}\left(Y_{\Delta ee},Y_{\Delta \mu\mu},Y_{\Delta\tau\tau}\right)\,,\\
\vec Y_\Delta=&\left(Y_{\Delta ee},Y_{\Delta \mu\mu},Y_{\Delta\tau\tau}\right)^t\,,\\
Y_\Delta^{\rm wo}=&Y_\Delta-Y_\Delta^{\rm diag}+{\rm diag}\left(A\, \vec Y_\Delta \right)\,,
\end{align}
\end{subequations}
where the superscript $t$ indicates a transposition.
The matrix components of the washout rate are given by
\begin{align}
{\cal B}^{Ni,k}_{\ell ab}=Y^\dagger_{ak}Y_{kb}\frac{3}{2^{7/2}\pi^{5/2}}
\left(\frac{M_{Nk}}{M_{Ni}}\right)^\frac52\frac{a_{\rm R}}{M_{Ni}}
z_i^\frac52{\rm e}^{-z_i\frac{M_{Nk}}{M_{Ni}}}\,.
\end{align}
Putting together the washout terms induced by lepton and by Higgs charge densities,
we eventually obtain for the washout term
\begin{align}
\bar W_\Delta[Y_\Delta]=\sum\limits_k \left(\frac12\{{\cal B}^{Ni,k}_{\ell},Y^{\rm wo}_\Delta\}+\frac 12{\cal B}^{Ni,k}_{\ell}(1,1,1)^t [C_\phi\vec Y_\Delta] \right)\,.
\end{align}
The factor of $1/2$ in front of the Higgs-induced term can be understood when noting
that $q_\ell=\mu_\ell T^2/6$, whereas $q_\phi=\mu_\phi T^2/3$, where
$\mu_{\ell,\phi}$ are chemical potentials and the factor two is due to
the difference between Fermi and Bose statistics.

\section{Parametric Surveys}
\label{sec:surveys}

\subsection{Parametrisation of the Yukawa Couplings}

Taking diagonal matrices for the sterile neutrino masses $M_N$, the neutrino
sector of the model~(\ref{Lagrangian}) yet encompasses $18$ parameters:
$3$ sterile neutrino masses and $15$ parameters in the Yukawa coupling $Y$. (While
the complex $3\times3$ matrix $Y$ has eighteen degrees of freedom, three
of these can be absorbed by phase rotations of the SM leptons $\ell$.)
This large number of parameters is a typical obstacle to comprehensive studies
of the parameter space in type-I seesaw models.

The Casas-Ibarra parametrisation~\cite{Casas:2001sr} facilitates to impose the observational
constraints from neutrino oscillations by rearranging the Lagrangian parameters
into low and high energy categories. The nine high energy parameters are given by
$M_{N1,2,3}$ as well as three complex angles $\varrho_{12}$,
$\varrho_{13}$, $\varrho_{23}$, in terms of which one defines the
complex orthogonal matrix
\begin{align}
{\cal R}=
\left(
\begin{array}{ccc}
{\rm c}_{12}{\rm c}_{13} & {\rm c}_{13}{\rm s}_{12} & {\rm s}_{13}\\
-{\rm c}_{23}{\rm s}_{12}-{\rm c}_{12}{\rm s}_{23}{\rm s}_{13} & {\rm c}_{23}{\rm s}_{12}-{\rm s}_{12}{\rm s}_{23}{\rm s}_{13} & {\rm c}_{13}{\rm s}_{23}\\
{\rm s}_{12}{\rm s}_{23}-{\rm c}_{12}{\rm c}_{23}{\rm s}_{13} & -{\rm c}_{12}{\rm s}_{23}-{\rm c}_{23}{\rm s}_{12}{\rm s}_{13} & {\rm c}_{13}{\rm c}_{23}
\end{array}
\right)\,,
\end{align}
where ${\rm s}_{ij}=\sin\varrho_{ij}$ and ${\rm c}_{ij}=\cos\varrho_{ij}$.

The nine low-energy parameters are given in terms of the
diagonal mass-matrix of the active neutrinos
$m_\nu={\rm diag}(m_1,m_2,m_3)$ and the six real angles and phases of the
PMNS matrix
\begin{align}
U_\nu=V^{(23)}U_\delta V^{(13)}U_{-\delta}V^{(12)}{\rm diag}({\rm e}^{{\rm i}\alpha_1/2},{\rm e}^{{\rm i}\alpha/2},1)\,,
\end{align}
where
\begin{align}
V^{(12)}=&
\left(
\begin{array}{ccc}
\cos\vartheta_{12} & \sin\vartheta_{12} & 0\\
-\sin\vartheta_{12} & \cos\vartheta_{12} & 0\\
0 & 0 & 1
\end{array}
\right)\,,
V^{(13)}=
\left(
\begin{array}{ccc}
\cos\vartheta_{13} & 0 &\sin\vartheta_{13}\\
0 & 1 & 0\\
-\sin\vartheta_{13} & 0 &\cos\vartheta_{13}\\
\end{array}
\right)\,,
\\\notag
V^{(23)}=&
\left(
\begin{array}{ccc}
1 & 0 & 0\\
0 & \cos\vartheta_{23}  &\sin\vartheta_{23}\\
0 & -\sin\vartheta_{23} &\cos\vartheta_{23}\\
\end{array}
\right)\,,
\end{align}
and
$U_{\pm\delta}={\rm diag}({\rm e}^{\mp{\rm i}\delta/2},1,{\rm e}^{\pm {\rm i}\delta/2})$.
In terms of this parametrisation, the Yukawa couplings of the sterile neutrinos
are obtained as
\begin{align}
\label{Casas:Ibarra}
Y^\dagger=U_\nu\sqrt{m_\nu}{\cal R}\sqrt{M_N}\frac{\sqrt{2}}{v}\,.
\end{align}

A considerable, yet generic simplification occurs when one of the three sterile
neutrinos decouples, say $N_3$ for definiteness. This can happen when the
Yukawa couplings $Y_{3a}$ are very small, when $M_3$ is very large or when
we only assume the existence of two sterile neutrinos to start with.
Note that such a configuration requires one of the
light neutrinos to be massless. If we therefore take $m_1=0$, we imply that
\begin{align}
\label{rhoangles:decoup}
\varrho_{23}=0\,\quad\varrho_{13}=\pi/2\,.
\end{align}
Moreover, the Yukawa couplings as given by Eq.~(\ref{Casas:Ibarra}) then
turn out to be independent of $\alpha_1$, as an immediate consequence of
$m_1=0$.

Altogether, in the decoupling scenario, there are $11$ Lagrangian parameters ($9$ parameters in
the Yukawa couplings $Y$ after rephasings and two Majorana masses for
the sterile neutrinos). These decompose into $4$ high energy parameters
($M_{N1,2}$ and $\varrho_{12}$) and  $7$ low-energy parameters
($m_2$, $m_3$, three angles $\vartheta_{ij}$ and the two phases $\delta$
and $\alpha_2$). Out of the latter,
$5$ have been measured experimentally ($\Delta m^2$, $\delta m^2\approx m_2^2$ and
the three PMNS mixing angles $\vartheta_{ij}$). The free parameters of the model
are therefore $M_{N1,2}$, $\varrho$, $\alpha$ and $\delta$, while for the PMNS mixing angles
in our numerical examples,
we choose
$\sin\vartheta_{12}=0.55$, $\sin\vartheta_{23}=0.63$ and $\sin\vartheta_{13}=0.16$,
which are close to the best-fit values determined by current observations~\cite{Fogli:2012ua,Tortola:2012te}.

\subsection{The Parameter Space in the Decoupling Scenario}

The production rates of the $N_i$, the washout rates of the
asymmetries as well as the $CP$ cuts entering the production rates
of the lepton asymmetries that are presented in Section~\ref{sec:freezeout}
apply all to the non-relativistic regime, {\it i.e.} when
$M_{Ni}\gg T$ for all $N_i$ that are involved in a certain rate.
In order to employ these results consistently, we should then
avoid situations when at times relevant for the freeze-out value
of the asymmetry relativistic $N_i$ are present.
For the purpose of the present analysis, we therefore choose
RHN masses that are not hierarchical, while not necessarily degenerate.
Besides, for the source from mixing SM leptons, such parametric
configurations are also favoured by
the fact that the asymmetry from the decay of the lightest of the
$N_i$ is exponentially suppressed in the
case of hierarchical $M_i$, {\it cf.} Eq.~(\ref{Sbar}) above
and Figure~\ref{fig:varyM2} below.
In future work, it may be of interest
though to consider relativistic RHNs when the asymmetry does not result from
the decay of the lightest RHN, as certain flavour correlations may generically
survive the washout from the lighter RHNs~\cite{DiBari:2005st,Antusch:2010ms,Garbrecht:2014bfa}.

Now, as we observe below, the dependence of the final asymmetry on the
relative size of $M_1$ and $M_2$ turns out to be mild ({\it cf.} Figure~\ref{fig:varyM2}).
Besides, given the relation~(\ref{Casas:Ibarra}) and the Boltzmann equations
from Section~\ref{sec:Boltzmann}, we see that the value of the freeze-out
asymmetry scales proportional to the $M_{Ni}$ when keeping the mass ratios
fixed.
Therefore,
a scan over the four-dimensional parameter space defined by $\varrho_{12}$,
$\delta$ and $\alpha_2$ yields comprehensive information on the model
in the decoupling scenario [with $\varrho_{23}$ and $\varrho_{13}$
as in Eqs.~(\ref{rhoangles:decoup})], given the constraints mentioned above.
For the purpose of the scan, we choose the masses of the two
RHNs as these are given in Table~\ref{table:3flav:opti}. The remaining values
specified in Table~\ref{table:3flav:opti} correspond to the point that we
find in parameter
space for which the maximal asymmetry occurs. We use the flavour approximations
as specified for Regime~B~1 in Section~\ref{sec:regimes}.

\begin{table}
\begin{center}
\begin{tabular}{|r|l||r|l||r|l|}
\hline
$M_{N1}$ & $3.67\times 10^9\,{\rm GeV}$ &$m_1$& $0\,{\rm meV}$& $\alpha_2$ & $1.7$\\
$M_{N2}$ & $3.2\times 10^9\,{\rm GeV}$ &$m_2$& $8.7\,{\rm meV}$& $\delta$ & $-0.4$\\
&&$m_3$&$49\,{\rm meV}$&$\varrho_{12}$ & $0.02+0.31{\rm i}$\\
\hline
\end{tabular}
\end{center}
\caption{\label{table:3flav:opti}Set of parameters that yields the largest asymmetry for
Leptogenesis from mixing lepton doublets in the decoupling scenario
(effectively two RHNs only).}
\end{table}

\begin{figure}[t!]
\begin{center}
\begin{tabular}{cc}
\hspace{-.4cm}
\epsfig{file=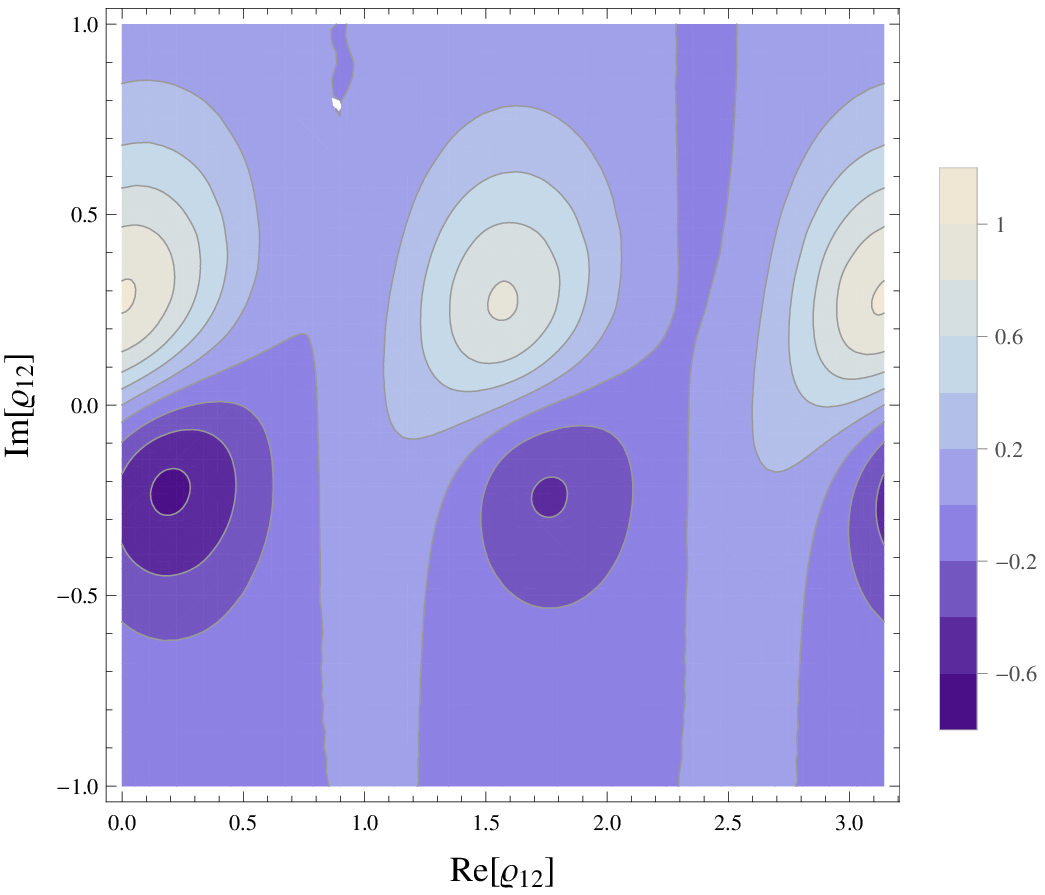,width=7.9cm}
&
\epsfig{file=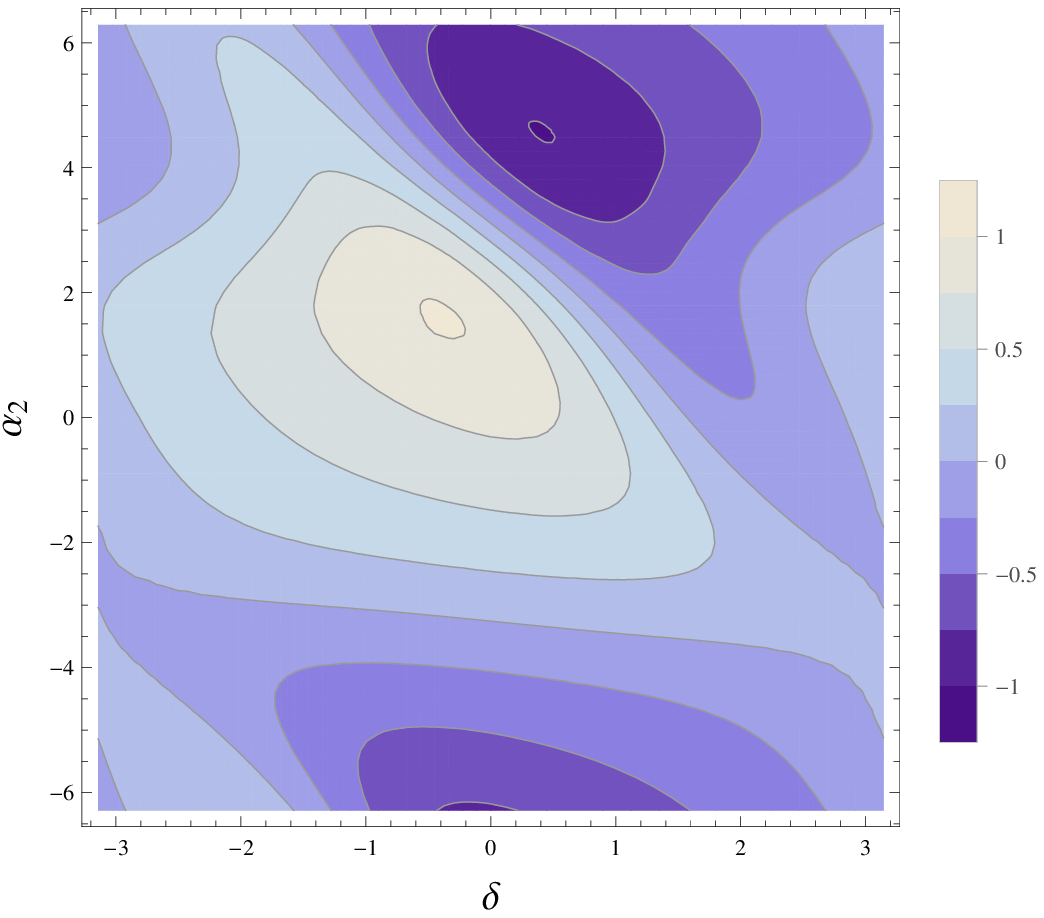,width=7.6cm}
\\
(A)&(B)
\end{tabular}
\end{center}
\vskip-.4cm
\caption{
\label{fig:optiscan}
Asymmetry $Y/Y_{\rm obs}$ in the $\varrho_{12}$ plane~(A) and in the $\delta$-$\alpha_2$ plane~(B),
with the remaining parameters as specified in Table~\ref{table:3flav:opti}.
}
\end{figure}

In Figure~\ref{fig:optiscan}, we show the freeze-out asymmetry
\begin{align}
Y=\sum\limits_{a,i} Y^{Ni}_{\Delta{aa}}(z_i\to \infty)
\end{align}
normalised to the observed
baryon-minus-lepton asymmetry~\cite{Hinshaw:2012aka,Ade:2013zuv}
\begin{align}
Y_{\rm obs}=\left(\frac{28}{79}\right)^{-1}\times8.6\times10^{-11}\,,
\end{align}
where the first factor accounts for the conversion to the final baryon asymmetry via
sphalerons~\cite{Harvey:1990qw}.
We vary parameters in the planes $\varrho_{12}$ and $\delta$ {\it vs.} $\alpha_2$, where we fix
the remaining parameters as in Table~\ref{table:3flav:opti}. The alignment of some of
the contours along $\delta+\frac{\alpha}{2}={\rm const.}$
in Figure~\ref{table:3flav:opti}(B) can be attributed to a constant washout
of the $e$-flavour, since we find for the Yukawa couplings that~({\it cf.} Refs.~\cite{Shuve:2014zua,Asaka:2011pb})
\begin{subequations}
\begin{align}
Y_{1e}=&\frac{\sqrt{2M_1}}{v}\left(-{\rm e}^{{\rm i}\frac{\alpha}{2}}\sqrt{m_2}\cos\vartheta_{13}\sin\vartheta_{12}\sin\varrho_{12}
-{\rm e}^{-{\rm i}\delta}\sqrt{m_3}\sin\vartheta_{13}\cos\varrho_{12}\right)\,,\\
Y_{2e}=&\frac{\sqrt{2M_2}}{v}\left({\rm e}^{{\rm i}\frac{\alpha}{2}}\sqrt{m_2}\cos\vartheta_{13}\sin\vartheta_{12}\cos\varrho_{12}
-{\rm e}^{{\rm i}\delta}\sqrt{m_3}\sin\vartheta_{13}\sin\varrho_{12}\right)\,.
\end{align}
\end{subequations}

\begin{figure}[t!]
\begin{center}
\begin{tabular}{cc}
\hspace{-.4cm}
\epsfig{file=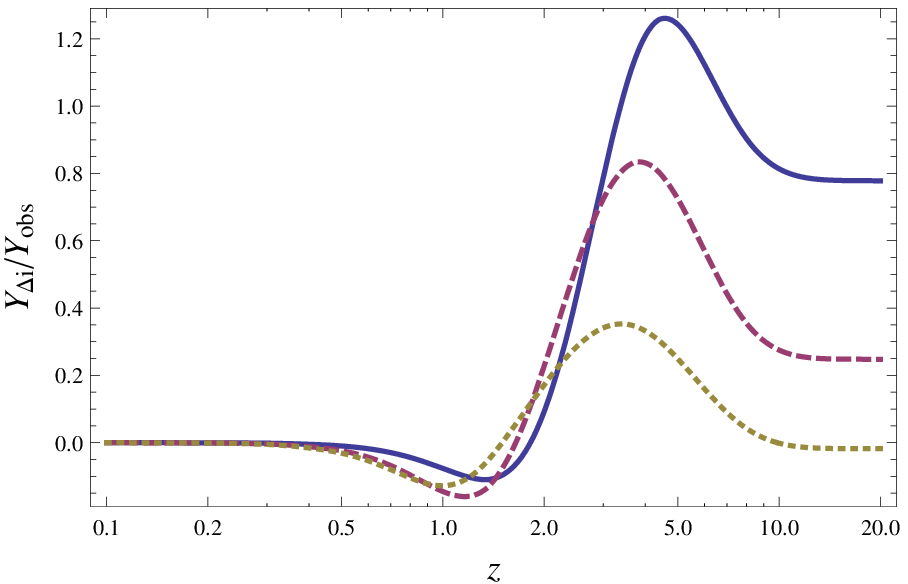,width=7.5cm}
&
\epsfig{file=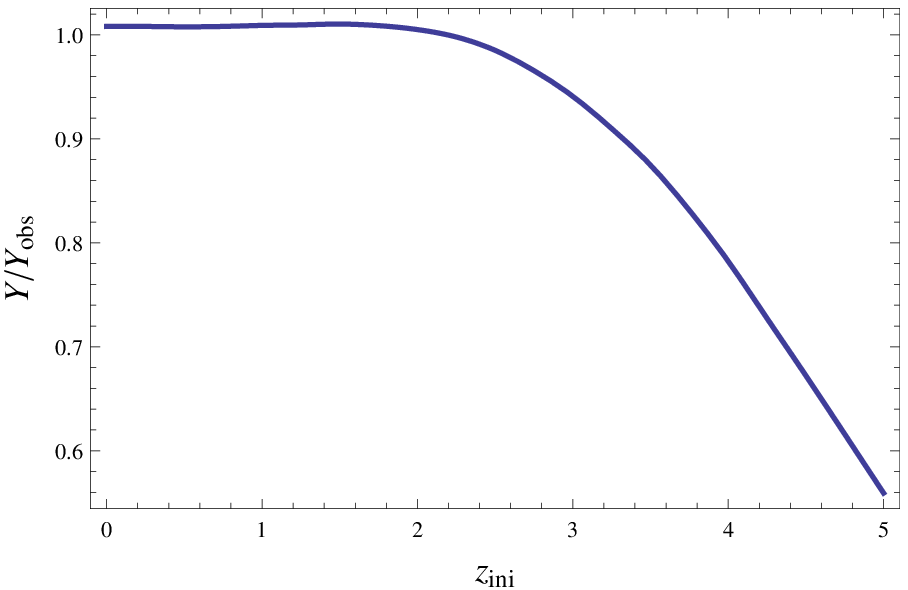,width=7.5cm}
\\
(A)&(B)
\end{tabular}
\end{center}
\vskip-.4cm
\caption{
\label{fig:zini}
In panel~(A), we show the evolution of the asymmetries
$Y_{\Delta i}$
with $i=e,\mu,\tau$ (solid blue, dashed red, dotted yellow) normalised to $Y_{\rm obs}$
over $z$. In panel~(B), we show the value of the freeze-out asymmetry (sum over three flavours)
over $z_{\rm ini}$. The parameters are given in Table~\ref{table:3flav:opti}.
}
\end{figure}

\begin{figure}[t!]
\begin{center}
\epsfig{file=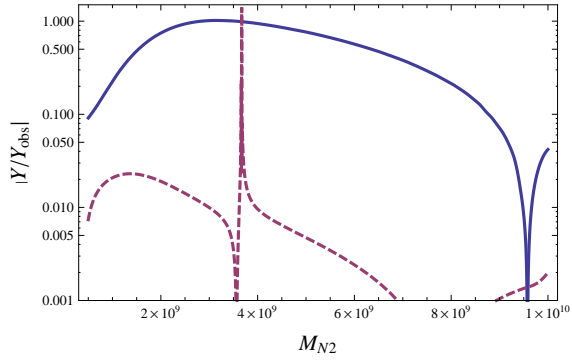,width=7.5cm}
\end{center}
\caption{
\label{fig:varyM2}
Dependence of the freeze-out asymmetry from lepton mixing on the parameter $M_{N2}$,
with the remaining parameters as specified in Table~\ref{table:3flav:opti} (solid blue).
For comparison, we show the asymmetry from standard Leptogenesis for the same parameters
(dashed red).
}
\end{figure}

Next, we validate the assumption of strong washout and non-relativistic RHNs by considering the
evolution of the individual flavour asymmetries $Y_{\Delta i}=Y_{\Delta ii}^{N1}+Y_{\Delta ii}^{N2}$
over the parameter $z=M_{N1}/T$, that
is commonly used as the time variable when studying Leptogenesis from massive
neutrinos. From Figure~\ref{fig:zini}~(A), we observe that, as it is typical for strong washout scenarios, the freeze-out value
of the asymmetry settles when $z\gtrsim 10$. In order to assess further the validity
of the non-relativistic approximation for the RHNs as well as in order to
determine the minimum value of the required reheat temperature, we
start the integration of the Boltzmann equations at some value $z=z_{\rm ini}$ with
vanishing asymmetries as boundary conditions (while for all other numerical results, we start
the integration at $z=z_{\rm ini}=0$). From Figure~\ref{fig:zini}~(B) we see
that the result changes by less than $10\%$ as long as $z_{\rm ini}\lsim 3$. This
independence of the details of the initial evolution of the asymmetries
is a typical feature of the strong washout regime. Given the value of $M_{N1}$
from Table~\ref{table:3flav:opti}, we may therefore conclude that the minimum
required reheat temperature $T_{\rm reh}$
in the decoupling scenario is $T_{\rm reh}\gsim1.2\times 10^9\,{\rm GeV}$. Due to the
order one uncertainties incurred through the estimate of $B_\ell^g$ and the momentum averaging leading to
$B_\ell^{\slashed{\rm fl}}$, this should be considered as coincident
with the bound of $T_{\rm reh}\gsim2\times 10^9\,{\rm GeV}$ for standard Leptogenesis~\cite{Davidson:2002qv,Buchmuller:2004nz,Davidson:2008bu}. However, the
present optimal (by the criterion of minimising the lower bound on $T_{\rm reh}$) point
given in Table~\ref{table:3flav:opti} is clearly distinct from the optimal parametric
configurations in standard Leptogenesis because here, we are in the strong washout regime,
while for the standard source, the lowest viable reheat temperatures occur in between the strong and
the weak washout regimes.

The analysis presented in Figure~\ref{fig:zini}~(B) also justifies the
use of the flavour approximations for Regime~B~1, valid for temperatures
roughly below $1.3\times 10^9\,{\rm GeV}$. While in fact, for $z\lsim 3$,
Our scenario falls into Regime~A~2, the final prediction for the asymmetry
should not be substantially affected by an inaccurate treatment of the flavour
effects at early times.

Finally, we vary $M_{N2}$ while keeping the remaining parameters
fixed as in Table~\ref{table:3flav:opti}. The resulting normalised freeze-out asymmetry is shown
in Figure~\ref{fig:varyM2}. We thus indeed verify that the ratio of $M_{N1}$ to $M_{N2}$ has no dramatic
influence on the freeze-out asymmetry as long it remains of order {\it one}. For comparison,
we also show in Figure~\ref{fig:varyM2} the asymmetry that arises for the same parameters from
standard Leptogenesis. While we clearly see the resonance for $M_{N2}\to M_{N1}$, away from
this narrow enhanced region, the result is small compared to the asymmetry arising from
lepton mixing. One should note however that there exist parametric configurations that
are more favourable for standard Leptogenesis, in particular when saturating the bound
on the reheat temperature from Refs.~\cite{Davidson:2002qv,Buchmuller:2004nz,Davidson:2008bu}.

\subsection{Three Sterile Neutrinos}
\label{sec:3RHNs}

Adding a third RHN $N_3$ implies that compared to the case with two RHNs only, the resulting asymmetry depends
in addition on $M_{N3}$, $\alpha_1$, $\varrho_{\rm 23}$,  $\varrho_{\rm 13}$ and the absolute mass scale of
the light neutrinos, {\it i.e.} there are seven extra parameters. This appears to prohibit a comprehensive
analysis of the parameter space in practice. Nonetheless, it is interesting to evaluate the asymmetries for
an example point, that would be consistent with a smaller reheat temperature. We discuss how an enhanced
asymmetry becomes possible even if generated at lower temperatures and how parameters need to be tweaked
in order to arrange for such a situation.

\begin{table}
\begin{center}
\begin{tabular}{|r|l||r|l||r|l||r|l|}
\hline
$M_{N1}$ & $3\times 10^8\,{\rm GeV}$ &$m_1$& $2.5\,{\rm meV}$& $\alpha_1$ & $-0.4$&$\varrho_{12}$&$0.01-0.05{\rm i}$\\
$M_{N2}$ & $4\times 10^8\,{\rm GeV}$ &$m_2$&  $9.1\,{\rm meV}$& $\alpha_2$ & $-0.2$&$\varrho_{23}$&$-0.19+0.19{\rm i}$\\
$M_{N3}$& $5\times 10^8\,{\rm GeV}$&$m_3$& $49\,{\rm meV}$&  $\delta$ & $-1.1$&$\varrho_{13}$&$2.1-3.0{\rm i}$\\
\hline
\end{tabular}
\end{center}
\caption{\label{table:3flav:3RHN}Parametric example point for the scenario
with three RHNs.}
\end{table}

\begin{figure}[t!]
\begin{center}
\begin{tabular}{cc}
\hspace{-.4cm}
\epsfig{file=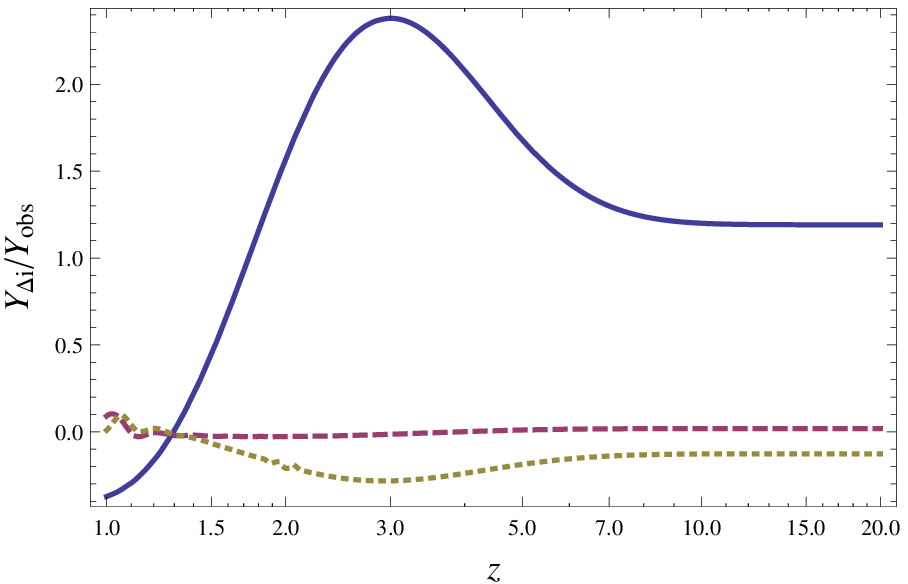,width=7.5cm}
&
\epsfig{file=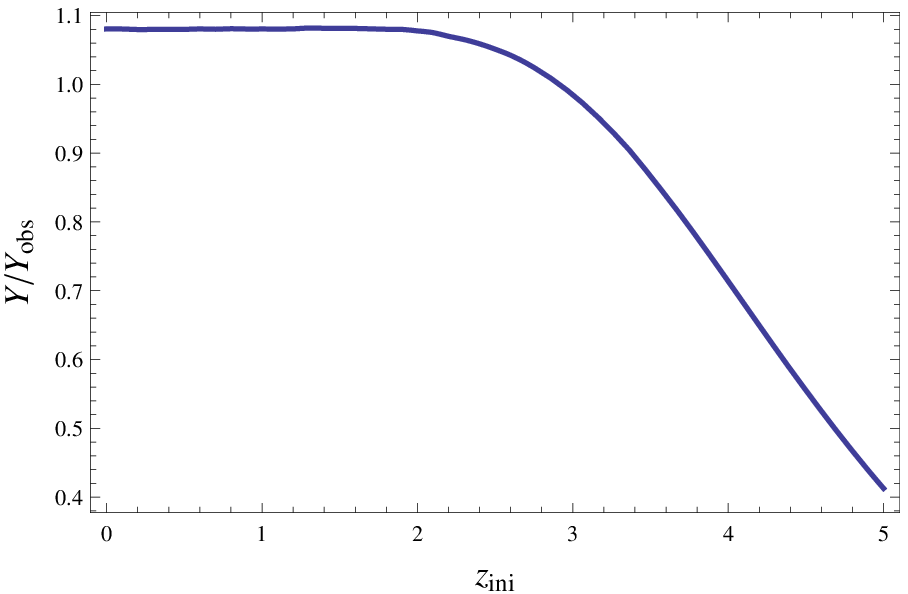,width=7.5cm}
\\
(A)&(B)
\end{tabular}
\end{center}
\vskip-.4cm
\caption{
\label{fig:zini:3RHN}
In panel~(A), we show the evolution of the asymmetries
$Y_{\Delta i}$
with $i=e,\mu,\tau$ (solid blue, dashed red, dotted yellow) normalised to $Y_{\rm obs}$
over $z$. In panel~(B), we show the value of the freeze-out asymmetry (sum over three flavours)
over $z_{\rm ini}$. The parameters are given in Table~\ref{table:3flav:3RHN}.
}
\end{figure}

In Table~\ref{table:3flav:3RHN}, we present the point in parameter space for which we determine the freeze-out
lepton asymmetry, where it can be seen from Figure~\ref{fig:zini:3RHN} that an asymmetry in accordance with
the observed value is obtained. Moreover, as exhibited in Figure~\ref{fig:zini:3RHN}(A),
the final asymmetry is dominated by the $e$-flavour. This can be understood by an inspection of the matrix of Yukawa couplings that satisfies $|Y_{ie}|\ll|Y_{i\mu}|,|Y_{i\tau}|$, such that there is
a substantially smaller washout rate for $\ell_e$ than for
$\ell_\mu$ and $\ell_\tau$. On the other hand, larger $Y_{i\mu}$ and $Y_{i\tau}$ enhance the asymmetry
$q_{\ell ee}$, {\it cf.} Eq.~(\ref{source:q}).

Turning to the parameters in Table~\ref{table:3flav:3RHN}, we observe the large imaginary part
for $\varrho_{13}$, which implies that the Yukawa couplings have a larger magnitude than
for configurations with smaller imaginary parts of the $\varrho_{ij}$. This implies that there
is a cancellation in individual terms contributing to the masses of the light neutrinos in the
see-saw mechanism, which may be interpreted as parametric tuning. It is noteworthy that situations
with large couplings of $\ell_{\mu,\tau}$ and relatively small couplings of $\ell_e$ to the RHNs are
also favoured in scenarios of Leptogenesis where the $CP$-violating source arises from the oscillations
of relativistic RHNs~\cite{Akhmedov:1998qx,Drewes:2012ma,Shuve:2014zua,Garbrecht:2014bfa,Canetti:2014dka}, (the so-called ARS scenarios, after the authors of 
Ref.~\cite{Akhmedov:1998qx}).
We emphasise that however, the source from active lepton mixing is different from the source from
RHN oscillations, and while the favoured parametric configurations bear similarities in the pattern of
Yukawa couplings, for given masses of the RHNs, the main contributions to the
asymmetries are generated in both scenarios at very different temperatures, {\it cf.} Ref.~\cite{Garbrecht:2014bfa}.

\section{Conclusions}
\label{sec:conclusions}

In this work, we have investigated in some detail the possibility
of generating the baryon asymmetry of the Universe from
the mixing of lepton doublets within
the SM extended by two or three RHNs. For this purpose,
we have introduced a diagrammatic representation of the underlying mechanism, and
we have discussed the dynamics of the flavour correlations of SM leptons at
various temperatures. We then have performed a comprehensive parametric study in the
setup with two RHNs in the type-I see-saw mechanism. For the case with three RHNs, we
have identified a way to achieve lower reheat temperatures that are consistent with the
observed baryon asymmetry.

We find that baryogenesis from mixing lepton doublets is a generically
viable scenario in the type-I see-saw framework, provided
\begin{itemize}
\item
there are RHNs present in the mass range between $10^9\,{\rm GeV}$ and
$10^{11}\,{\rm GeV}$ ({\it cf.} the discussion of Section~\ref{sec:flavasymm} concerning
the upper bound and the numerical findings of Section~\ref{sec:surveys} regarding the lower
bound)
\item
and these RHNs are of the same mass-scale, while they do not need to be degenerate.
\end{itemize}
The lower mass bound on the RHNs and consequently on the reheat temperature
can be evaded through a certain alignment of the Yukawa couplings $Y$,
that allows these to be relatively large while the masses of the light active neutrinos remain
small, which is possible in the presence of three RHNs, {\it cf.} Section~\ref{sec:3RHNs}.

Methodically, the present calculation draws from formulations of
Leptogenesis in the CTP approach that have been applied to
the resonant regime~\cite{Garbrecht:2011aw}, to oscillations of relativistic RHNs~\cite{Drewes:2012ma}
as well as to the decoherence of active lepton flavours~\cite{Beneke:2010dz}.
To this end, we identify the quantities $B_\ell^{\slashed{\rm fl}}$ and
$B_\ell^g$ as the main contributors to the theoretical uncertainty.
In introducing these, we average over the lepton momentum-modes under the simplifying assumption
of identical
reaction rates. While such a procedure is common practice
in similar calculations for Leptogenesis from oscillations of RHNs
({\it cf. e.g.} Ref.~\cite{Asaka:2005pn,Drewes:2012ma,Asaka:2011wq}),
it would nonetheless be desirable to improve on
this approximation in the future by resolving the different reaction
rates for  each momentum mode.

It is interesting to observe that the minimal reheat temperatures for standard
Leptogenesis and for baryogenesis from mixing lepton doublets appear to coincide.
For couplings as in the SM, the term $(h_{aa}^2+h_{bb}^2)B_\ell^{\slashed{\rm fl}}B_\ell^g$
in the enhancement factor, Eq.~(\ref{Qellab}), numerically dominates in the denominator.
Smaller gauge couplings would therefore lead to a larger asymmetry.
On the other hand,
the size of the gauge couplings does not have 
a leading influence on the asymmetry for standard Leptogenesis~\cite{Davidson:2002qv,Buchmuller:2004nz,Davidson:2008bu}. Therefore,
the similar bound on the reheat temperatures can be attributed to a
parametric coincidence.

We also note that
the analysis in the present study is valid for the strong washout regime, {\it i.e.} the situation
where
the RHNs can be approximated as non-relativistic during the creation of
the asymmetry. It would be interesting to relax this assumption (which should be possible
for at least one of the RHNs when three or more RHNs are present altogether) because
one may then anticipate substantially larger deviations from equilibrium. In that
case, a calculation would however not enjoy the considerable simplifications that arise
from treating the RHNs as non-relativistic.

While we have considered here the somewhat minimal framework of the SM augmented
by RHNs, our analysis implies that
new gauged particles that share the same quantum numbers and
that are nearly degenerate or effectively become degenerate at higher
temperatures are generic candidates for being involved in creating
the matter-antimatter asymmetry.
This opens new prospects for
scenarios of baryogenesis from out-of-equilibrium reactions in the
expanding Universe.

\subsection*{Acknowledgements}
We acknowledge
support by the Gottfried Wilhelm Leibniz programme
of the Deutsche Forschungsgemeinschaft and by
the DFG cluster of excellence ‘Origin and Structure of the Universe’.

\end{document}